%% file: main.tex
\documentclass[manuscript]{acmart}

\usepackage[final]{showlabels}
\usepackage{booktabs}
\usepackage{anyfontsize}
\usepackage{amsfonts}
\usepackage{amsmath}
\usepackage{graphicx}
\usepackage{balance}
\usepackage[T1]{fontenc}
\usepackage{subfigure}
\usepackage{algorithm2e}
\usepackage{amsthm}
\usepackage{hyperref}
\usepackage{xcolor}
\usepackage{pifont}
\usepackage{multirow}
\usepackage{array}
\usepackage[inline]{enumitem}
\usepackage{bbding}
\usepackage{wasysym}
\usepackage[table]{xcolor}

\usepackage[skip=0.5\baselineskip]{caption}

\hypersetup{
    colorlinks=false,
    linkcolor={red!20!black},
    citecolor={green!20!black},
    urlcolor={blue!20!black}
}

\usepackage{listings}

\lstdefinelanguage{SPARQL}{
  morekeywords={
    SELECT, CONSTRUCT, ASK, DESCRIBE,
    WHERE, FILTER, OPTIONAL, GRAPH,
    UNION, PREFIX, BASE, LIMIT, OFFSET, ORDER, BY, ASC, DESC,
    DISTINCT, REDUCED, FROM, NAMED
  },
  sensitive=true,
  morecomment=[l]{\#},
  morestring=[b]",
}

\setcopyright{cc}
\setcctype{by}
\acmJournal{TWEB}
\acmYear{2026} \acmVolume{1} \acmNumber{1} \acmArticle{}
\acmMonth{1} \acmDOI{10.1145/3847195}

\newcounter{todocounter}
\newcommand{\alltodos}{}

\makeatletter
\newcommand{\TODO}[1]{
  \refstepcounter{todocounter}
  \edef\@todoid{\number\value{todocounter}}
  \phantomsection
  \label{todo:\@todoid}
  \textcolor{red}{\textbf{[TODO: #1]}}
  \protected@edef\@tempa{%
    \noexpand\gappto\noexpand\alltodos{%
      \noexpand\item
      \noexpand\hyperref[todo:\@todoid]{\unexpanded{#1} (p.~\noexpand\pageref{todo:\@todoid})}
    }
  }
  \@tempa
}
\makeatother

\title[SPARQL-LLM]{SPARQL-LLM: Real-Time SPARQL Query Generation from Natural Language Questions 
}

\author{Panayiotis Smeros}\authornote{Equal contribution}
\affiliation{%
  \institution{SIB Swiss Institute of Bioinformatics}
  \city{Lausanne}
  \country{Switzerland}}
\email{panayiotis.smeros@sib.swiss}

\author{Vincent Emonet}\authornotemark[1]
\affiliation{%
  \institution{SIB Swiss Institute of Bioinformatics}
  \city{Lausanne}
  \country{Switzerland}}
\email{vincent.emonet@sib.swiss}

\author{Ruijie Wang}
\affiliation{%
  \institution{SIB Swiss Institute of Bioinformatics and University of Zurich}
  \city{Zurich}
  \country{Switzerland}}
\email{ruijie.wang@sib.swiss}

\author{Ana-Claudia Sima}
\affiliation{%
  \institution{SIB Swiss Institute of Bioinformatics}
  \city{Zurich}
  \country{Switzerland}}
\email{ana-claudia.sima@sib.swiss}

\author{Tarcisio Mendes de Farias}
\affiliation{%
  \institution{SIB Swiss Institute of Bioinformatics and University of Lausanne}
  \city{Lausanne}
  \country{Switzerland}}
\email{tarcisio.mendes@sib.swiss}
\renewcommand{\shortauthors}{Smeros et al.}

\begin{document}
\input{00-abstract}
\maketitle
\input{01-introduction}

\input{02-related_work}

\input{03-overview}
\input{04-evaluation}
\input{05-conclusion}


\clearpage
\balance
\bibliographystyle{ACM-Reference-Format}
\bibliography{references}




\end{document}

%% file: 00-abstract.tex

\begin{abstract}
The advent of large language models is contributing to the emergence of novel approaches that promise to better tackle the challenge of generating structured queries, such as SPARQL queries, from natural language. However, these new approaches mostly focus on response accuracy while ignoring other evaluation criteria, such as runtime and cost to generate SPARQL queries. Consequently, they are often not production-ready or easy to deploy over real-world knowledge graphs with good accuracy. To mitigate these issues, in this paper, we describe and systematically evaluate SPARQL-LLM, an open-source and triplestore-agnostic approach, powered by lightweight metadata, that generates SPARQL queries from natural language text.
First, we describe its architecture, which consists of dedicated components for metadata indexing, prompt building, and query generation and execution.
Then, we evaluate it based on a state-of-the-art challenge with multilingual questions, and a collection of questions from three of the most prevalent knowledge graphs within the field of bioinformatics.
Our results demonstrate
a substantial improvement of up to $59\%$ in F1 score over the second-best system participating in the challenge,
adaptability to high-resource languages such as English, Spanish, and German, as well as ability to form complex bioinformatics queries.
Furthermore, our results show that our system is up to $27\times$ faster than the second-best system participating in the challenge, while costing a maximum of $\$0.01$ per question, making it suitable for real-time, low-cost text-to-SPARQL applications.
SPARQL-LLM is publicly released as an open-source project at \url{https://github.com/sib-swiss/sparql-llm} and is currently deployed over real-world decentralized knowledge graphs at \url{https://www.expasy.org/chat}.
\end{abstract}

%

\ccsdesc[500]{Information systems~Question answering}
\ccsdesc[500]{Information systems~Information extraction}
\ccsdesc[500]{Information systems~Query suggestion}
\ccsdesc[500]{Information systems~Query reformulation}
\ccsdesc[300]{Information systems~Wrappers (data mining)}

\keywords{Large Language Models, SPARQL, TEXT2SPARQL, Bioinformatics, Knowledge Graphs} 

%% file: 01-introduction.tex

\section{Introduction}\label{sec:introduction}

In modern web-based ecosystems, accessing distributed data remains a critical challenge for uncovering valuable knowledge.
With the abundance of available data, and given its heterogeneity in formats, schemas, and access methods, as well as the increasing need for compliance with the FAIR principles \cite{wilkinson2016fair}, representing this information as Knowledge Graphs (KGs) has emerged as one of the few feasible solutions.
However, formulating queries using the SPARQL query language over these KGs has been shown to be an ``absurdly difficult'' task \cite{DBLP:journals/bmcbi/McCarthyVW12}, requiring familiarity with:
\begin{enumerate*}[label=\roman*)]
\item the syntax of the query language, 
\item the specifications and potential limitations of the access methods (e.g., the triplestore query engines),
\item the schema/ontology information of the underlying data, and
\item the potential interconnections across this data.
\end{enumerate*}
The task becomes even more challenging in specialized domains, where the queries are particularly complex, and even domain-experts often spend considerable time formulating, testing, and refining those queries.

Over the years, there has been a growing interest in Knowledge Graph Question Answering (KGQA) systems, which aim to bridge the gap between natural language questions and KGs by automatically translating these questions into appropriate SPARQL queries that can be posed over the underlying KGs \cite{DBLP:conf/ijcai/LanHJ0ZW21, diefenbach2018core}.
By eliminating the need for manual query crafting, KGQA systems can broaden access to KGs and accelerate scientific data discovery, especially in domains such as bioinformatics, where data volume and schema complexity evolve rapidly.

The latest breakthroughs in Large Language Models (LLMs) have provided a promising foundation for next-generation KGQA systems.
Their strong linguistic and reasoning capabilities enable them to translate natural language directly into SPARQL, with ``minimal prompting and without fine-tuning''
\cite{dabramo-etal-2025-investigating}.
Recent LLM-based KGQA approaches commonly incorporate the following two methods \cite{DBLP:conf/emnlp/MaCWKW25}: 
\begin{enumerate*}[label=\roman*)]
\item retrieval of contextual information, such as example queries or schema fragments, that can guide query construction, and
\item agent-based reasoning, where the LLM iteratively calls functions to explore the knowledge graph.
\end{enumerate*}
Nevertheless, existing KGQA systems frequently perform poorly in formulating queries for large, rapidly evolving, and highly specialized KGs, such as those encountered in the field of bioinformatics, which has been shown to be a demanding task even for domain-experts \cite{DBLP:conf/esws/SimaF23, sib2024sib}.


In this paper, we describe \emph{SPARQL-LLM}, an open-source KGQA system that generates SPARQL queries from natural language questions, leveraging lightweight metadata extracted directly from SPARQL endpoints.
The main contributions of the paper are described as follows:

\begin{itemize}
\item 
We propose a hybrid approach that combines retrieval-based KGQA, for leveraging prior knowledge to generate SPARQL queries, with the iterative refinement capabilities of agent-based KGQA, for validating these queries.
\item We design a modular architecture comprising dedicated components for metadata indexing, prompt construction, and query generation and execution, enabling efficient and production-ready KGQA applications.
\item We conduct a systematic evaluation using both a multilingual state-of-the-art challenge and prevalent real-world KGs within the field of bioinformatics, demonstrating:
\begin{itemize}
\item
a substantial improvement of up to $59\%$ in F1 score over the second-best system participating in the challenge;
\item adaptability to high-resource languages such as English, Spanish, and German; and

\item handling complex bioinformatics queries.
\end{itemize}

\item We showcase that SPARQL-LLM also achieves: 
\begin{itemize}
\item the lowest latency, being up to $27\times$ faster than the second-best system participating in the challenge; and
\item minimal operational cost, requiring only as little as $\$0.01$ per question.
\end{itemize}

\item We publicly release SPARQL-LLM as an open-source project to support transparency, reproducibility, and extensibility by the research community at \url{https://github.com/sib-swiss/sparql-llm}.

\item  We deploy SPARQL-LLM in production over real-world decentralized KGs in the field of bioinformatics, making them accessible to both SPARQL experts and non-experts at \url{https://www.expasy.org/chat}.
\end{itemize}

We highlight that this paper extends our prior preliminary work \cite{emonet2024llmbasedsparqlquerygeneration}, providing:
\begin{enumerate*}[label=\roman*)]
\item a thorough overview of related work,
\item a detailed description of the overall architecture and the individual components of our system, including a completely new component (\emph{SPARQL Execution}) and modules that were not part of the system, and most notably,
\item a formal evaluation, which in our prior work was based solely on a limited, non-standardized test set.
\end{enumerate*}

%% file: 02-related_work.tex
\section{Related Work}\label{sec:related_work}
Our related work consists of two categories of systems: 
\begin{enumerate*}[label=\roman*)]
\item agent-based systems for iteratively exploring the underlying knowledge graph, and
\item retrieval-based systems for obtaining contextual information that guides query construction.
\end{enumerate*}

MetaboT \cite{bekbergenova2025metabot}, ARUQULA \cite{Brei2025}, GRASP \cite{walter2025grasp}, 
Graf von Data \cite{Dorsch2025}, mKGQAgent \cite{Perevalov2025}, and PYTHIA \cite{10.1145/3770855.3817968}
are representative systems where an LLM is paired with a toolbox and iteratively reasons about which actions to take in order to build a SPARQL query. 
These systems typically expose generic functions, such as \texttt{search\_entity}
and \texttt{search\_examples}, which can be called repeatedly to discover relevant knowledge. 
%
While such approaches are attractive in principle, since sufficiently expressive tools allow them to adapt to arbitrary queries, they also face critical limitations. 
First, efficiency remains a challenge; each new question often requires rediscovering schema patterns from scratch, resulting in long runtimes (e.g., dozens of tool calls) and high costs due to token usage. 
Second, complex query patterns, such as property paths across multiple classes, 
remain difficult to handle; agents may fail to converge or require many iterations. 

On the other hand, 
T2SPARQL \cite{Berezin2025} and IRIS \cite{DBLP:conf/llms4ol/LatipovHM25}
are representative systems where an LLM is paired with a retrieval-based context construction, injecting data shapes and ontology fragments into the LLM prompt.
Furthermore, 
StarCoder \cite{Soru2025} is a representative system utilizing fine-tuned autoregressive models for query generation.

SPARQL-LLM follows a hybrid approach, aiming for real-time query generation with minimal interactions with the LLM.
Our approach combines retrieval-based KGQA, for leveraging prior knowledge to generate SPARQL queries, with the iterative refinement capabilities of agent-based KGQA, for validating these queries.


In terms of evaluation, existing work mostly targets encyclopedic questions over knowledge graphs such as DBpedia
and Wikidata.
For example, Liu et al.~\cite{DBLP:conf/emnlp/LiuSTXZL24} leveraged the Wikidata Request a Query\footnote{\url{https://www.wikidata.org/wiki/Wikidata:Request_a_query}} forum to construct a challenging dataset called SPINACH, which includes complex natural language questions with expert-annotated SPARQL queries.
They demonstrate the limitations of recent knowledge graph question answering methods (such as ToG~\cite{DBLP:conf/iclr/SunXTW0GNSG24} and WikiSP~\cite{DBLP:conf/emnlp/XuLCPWSL23}) and propose an LLM-based agent that mimics expert's behavior to achieve superior performance in generating SPARQL queries.
In the scientific domain, there have been recent efforts to answer questions over scholarly knowledge graphs.
For example, Banerjee et al.~\cite{DBLP:conf/birws/BanerjeeAUB23} proposed DBLP-QuAD, which is a question answering dataset with 10,000 scholarly questions and SPARQL query annotations over the DBLP knowledge graph.
Based on DBLP-QuAD, Wang et al. developed NLQxform~\cite{DBLP:conf/semweb/0003ZRRB23}, a transformer-based model that translates scholarly questions into executable SPARQL queries.
To the best of our knowledge, SPARQL-LLM is the first approach evaluated on general-knowledge
and specialized-in-the-bioinformatics-domain questions, 
showcasing remarkable performance in both.

%% file: 03-overview.tex

\section{SPARQL-LLM Overview}\label{sec:system}
In this section, we provide an overview of our text-to-SPARQL approach, focusing on how we retrieve and/or generate the provided context, as well as on the overall architecture of our system.

\subsection{Provided Context}\label{subsec:provided_context}

We designed our text-to-SPARQL approach, inspired by how SPARQL-savvy users typically write queries. 
Given that SPARQL-LLM has access to a SPARQL endpoint, we leverage the following sources of information (ideally part of the endpoint documentation) to accurately craft SPARQL queries from natural language text:
\begin{enumerate}
    \item pairs of SPARQL query examples and their descriptions; and
    \item data-aware schema, potentially accompanied by frequency indicators.
\end{enumerate} 

We note that neither of these requirements is unrealistic.
Regarding the first requirement, in practice, many widely used SPARQL endpoints already provide curated query examples to help users understand how their data can be accessed,
including Wikidata Query Service\footnote{\url{https://query.wikidata.org}}, WikiPathways\footnote{\url{https://sparql.wikipathways.org}}, and dblp\footnote{\url{https://sparql.dblp.org}}. Such examples serve a role analogous to usage examples or API documentation in software systems: they make the structure, vocabulary, and intended access patterns of a data source more transparent, thereby lowering the barrier for users to formulate effective queries.

Regarding the second requirement, we utilize an open source VoID\footnote{\url{https://www.w3.org/TR/void}} generator to generate statistics (i.e., auto-generated metadata) over any arbitrary SPARQL endpoint.
We interchangeably utilize two triplestore-agnostic VoID generators that support large, billion-scale triplestores: 
\begin{enumerate*}[label=\roman*)]
\item a tool\footnote{\url{https://github.com/sib-swiss/void-generator}} that applies brute-force schema exploration, providing a complete metadata set with a high cost in terms of execution time, or
\item an internal tool of SPARQL-LLM, that provides a good trade-off between execution time and metadata set completeness, and is particularly useful for RDF datasets that often change and require constantly up-to-date VoID descriptions. 
\end{enumerate*}

Moreover, in our ablation study (\S~\ref{sec:system_analysis}), we analyze the impact of each component (i.e., examples and schema) independently, in order to estimate their individual contribution to the task of text-to-SPARQL translation.



\subsubsection{Query Examples}
SPARQL query examples play a crucial role in SPARQL-LLM.
In particular, we argue that query examples are most effective when they consist of real‑world, human‑crafted queries rather than self‑generated ones. This observation aligns with recent evidence showing that self‑generated procedural knowledge does not provide benefits on average, as models cannot reliably author the skills they later exploit \cite{li2026skillsbenchbenchmarkingagentskills}. The underlying reason is the substantial gap between user intent, typically expressed at a high level of abstraction (e.g., a natural‑language question), and the concrete formulation of a SPARQL query that correctly retrieves the answer. Consequently, grounding query examples in authentic end‑user perspectives is essential to accurately capture user intent and bridge this  gap.


In previous work \cite{10.1093/gigascience/giaf045}, we described a recommended format for representing such queries with minimal metadata, which makes SPARQL-LLM easily reusable across any knowledge graph that adheres to this metadata format. 
Furthermore, we introduced a collection of over $1,000$ question–query examples, distributed across different knowledge graphs in the field of bioinformatics.
The provenance of these examples is based on human curation, while they are available through a public repository\footnote{\url{https://github.com/sib-swiss/sparql-examples}}.
These examples are represented as an RDF graph using SHACL and schema.org vocabulary terms.
%


\subsubsection{Data-Aware Schema}
To leverage the data schema information, we retrieve the VoID descriptions from each endpoint, which detail the relationships between subject classes and object classes or datatypes via specific predicates. For example, the \textit{Protein} class is linked to the \textit{Gene} class through the \textit{encodedBy} predicate. 
This retrieved information allows us to generate simple, human-readable Shape Expressions (ShEx) for each class. Each object property references a list of classes rather than another shape, making each shape self-contained and interpretable on its own.
The generated ShEx representations are well-suited for use with an LLM, as they provide information about which predicates are available for a class, and the corresponding classes or datatypes to which those predicates point. For example, here is the shape generated for a \textit{Disease Annotation}:

\begin{verbatim}
shape:up_Disease_Annotation {
  a [ up:Disease_Annotation ] ;
  up:sequence [ up:Chain_Annotation up:Modified_Sequence ];
  rdfs:comment xsd:string ; up:disease IRI }
\end{verbatim}

We chose ShEx as the schema specification language as a practical design choice for our system;
ShEx is specifically designed to express schema constraints over RDF graphs and provides a concise representation that reduces the number of tokens processed by the LLMs, thereby lowering inference costs and potentially mitigating the “needle-in-a-haystack” effect observed in LLMs \cite{liu2024lost}. 
Recent work also suggests that more compact RDF syntaxes lead to better data extraction performance with language models \cite{ringwald202412}.
Although alternative languages such as OWL and SHACL could also be used, we leave a systematic comparison of their advantages and drawbacks in this context as future work.

\begin{figure*}[t]
  \centering
  \includegraphics[width=1\textwidth]{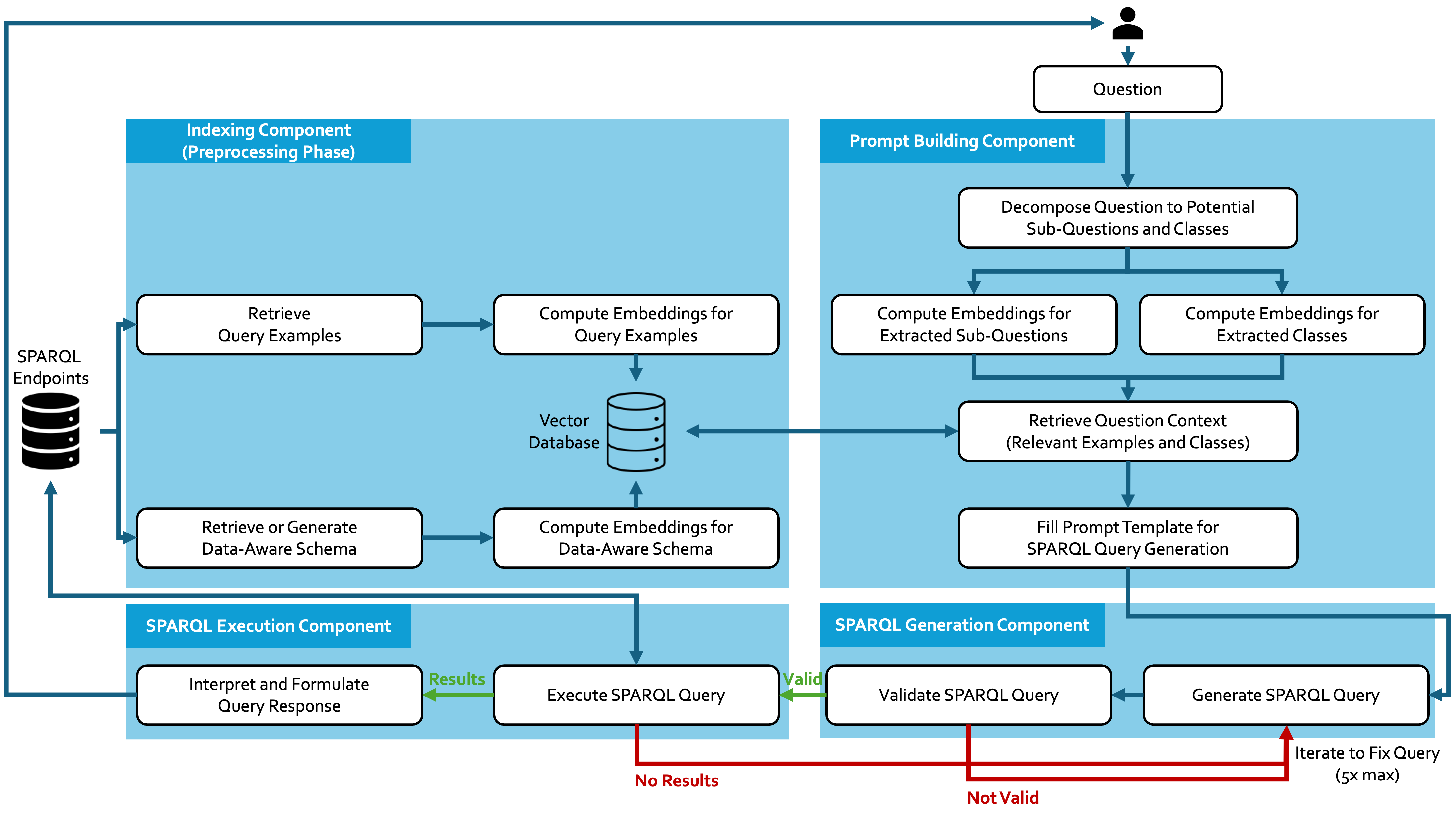}
  \caption{Architecture of SPARQL-LLM consisting of Indexing, Prompt Building, and SPARQL Generation and Execution components.}
  \Description{Diagram showing the architecture of the SPARQL-LLM system.}
  \label{fig:arch}
\end{figure*}

\subsection{System Architecture}

The proposed SPARQL-LLM system is illustrated in Figure~\ref{fig:arch}. The system takes a list of SPARQL endpoint URLs as input, where each endpoint is expected to include the minimal standardized metadata described above (i.e., example queries and VoID descriptions). This metadata is automatically retrieved and indexed upon initial deployment; we provide an online webpage\footnote{\url{https://sib-swiss.github.io/sparql-editor/check}} that allows users to check if a given endpoint contains the required metadata.

The overall data flow of the system is illustrated in Figure~\ref{fig:flow} and can be summarized as follows: 
\begin{enumerate*}[label=\roman*)]
\item decompose question by extracting potential sub-questions and identifying potential schema classes,
\item compute embeddings for the extracted sub-questions and identified classes, 
\item retrieve the relevant context (related examples and schema classes) using embeddings-based similarity search,
\item fill prompt template with the retrieved context for SPARQL query generation,
\item generate SPARQL query corresponding to the question,
\item validate and correct the query iteratively using schema information,
\item execute the validated query, and finally
\item interpret and present query response to the user.
\end{enumerate*}

\subsubsection{Indexing Component}\label{sec:indexing_component}

This component is responsible for indexing examples and schema classes that comprise the context provided to SPARQL-LLM (\S\ref{subsec:provided_context}).
Particularly, this component: 
\begin{enumerate*}[label=\roman*)]
\item retrieves or computes the context for each endpoint,
\item generates embeddings for this context (i.e., for each individual example and class of the endpoint), and
\item indexes these embeddings into a vector database.
\end{enumerate*}
%
This component is executed only during the initial deployment of the system (Figure~\ref{fig:arch}), enabling the subsequent semantic retrieval of the \emph{Prompt Building Component}.

%
%

\subsubsection{Prompt Building Component}
This component is responsible for processing a user question and building the contextualized prompt for the LLM.
Each question undergoes a decomposition phase using the LLM prompt described in Figure~\ref{fig:flow}, where: 
\begin{enumerate*}[label=\roman*)]
\item we split complex questions into standalone sub-questions, and
\item we extract high-level concepts that may correspond to classes present in the SPARQL endpoints.
\end{enumerate*}
We note that, the extracted sub-questions and high-level concepts are used solely to retrieve more relevant examples and classes from the vector database, which are combined into a single prompt.
Furthermore, if the question is already atomic, no decomposition is performed.
In the example shown in Figure~\ref{fig:flow}: 
\begin{enumerate*}[label=\roman*)]
\item we split the question \emph{``What is the HGNC symbol for the P68871 protein''} into the sub-questions: \emph{``Find protein''} and \emph{``HGNC symbol for protein''}, and
\item we extract the high-level concepts: \emph{``Protein''} and \emph{``Gene''}.
\end{enumerate*}

For each of
the extracted sub-questions and high-level concepts, we query the vector database to retrieve contextual information (i.e., relevant examples and classes) associated with the initial question. 
We follow a standard information retrieval methodology;
we project them into a shared embeddings space and compare them with the already indexed examples and classes using the cosine similarity metric. 
For each of the extracted sub-questions, we retrieve the $20$ most similar examples, while for each of the extracted high-level concepts, we retrieve the $15$ most similar classes.
Finally, we integrate this retrieved contextual information into the \emph{SPARQL Generation Prompt} (Figure~\ref{fig:flow}).

\subsubsection{SPARQL Generation Component}

This component is responsible for generating SPARQL queries, using the prompt built by the \emph{Prompt Building Component}, and interacting appropriately with an LLM.
%
We note that our approach is transparent to any underlying LLM; in fact, we evaluate the performance of several state-of-the-art LLMs in Section~\ref{sec:evaluation}.

After the generation phase, we validate and fix erroneous SPARQL queries using the provided schema information and potential error messages produced during query parsing and execution.
Aiming for real-time SPARQL generation, we omit full ontology reasoning and perform lightweight syntactic and structural checks on the generated query.
In more detail, we parse the generated query, extract its triple patterns, and check whether they are compatible with the provided schema.
We note that, if the provided schema contains entailed knowledge, the validator will also use this knowledge when validating a query; nevertheless, we cannot guarantee the semantic correctness of the query, as the validator does not perform the full range of entailment and consistency checks supported by an OWL reasoner.

In case of errors, the validator produces a list of human-readable error messages
(e.g., syntax, execution, and runtime errors reported by the SPARQL parser and query engine, and incompatibilities between the classes and predicates in the generated query and the provided schema).
These error messages are then incorporated into a new prompt and fed back to the LLM to iteratively refine the query. 
This targeted feedback loop helps constrain hallucinations and typically leads to an executable SPARQL query in up to three refinement iterations.

An illustrative example is shown in Figure~\ref{fig:flow}.
After analyzing the \emph{Generated Query}, our validator detects an error; the variable \texttt{?hgnc} with type \texttt{up:Resource} does not support the predicate \texttt{rdfs:label}.
Then, our system constructs a new prompt, describing the error and providing possible solutions to resolve it.
These solutions are derived from the indexed schema information (e.g., a valid predicate for the class \texttt{up:Resource} is \texttt{rdfs:comment}).
Finally, this prompt is fed back to the LLM, which responds with the \emph{Fixed Query}, where \texttt{rdfs:label} is replaced by \texttt{rdfs:comment}.




\subsubsection{SPARQL Execution Component}

\begin{figure*}[t]
  \centering
  \includegraphics[width=1\textwidth]{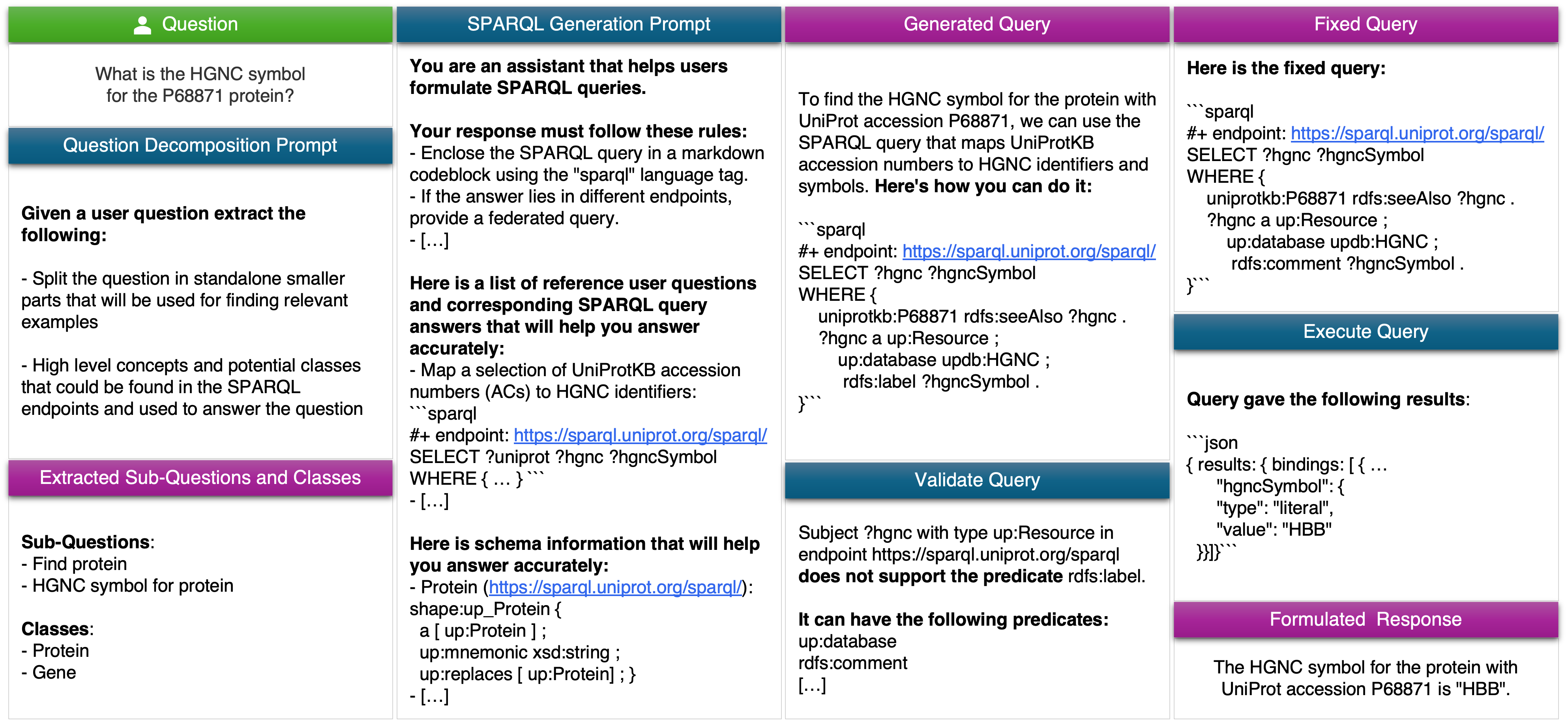}
  \caption{Data flow of SPARQL-LLM when resolving a complex bioinformatics question.}
  \Description{Diagram showing the data flow of the SPARQL-LLM system.}
  \label{fig:flow}
\end{figure*}

Once a query has been successfully generated and validated, it is executed against the corresponding SPARQL endpoint. 
The raw results of the execution are then fed back to the LLM, which interprets them in the context of the original user question. 
This allows the system not only to return the query output, but also to provide a concise, user-friendly explanation of the results (Figure~\ref{fig:flow}). 
In the case of federated queries, the system ensures that results from multiple endpoints are combined consistently before being forwarded to the LLM for interpretation.

%% file: 04-evaluation.tex

\section{Evaluation}\label{sec:evaluation}

In this section, we evaluate SPARQL-LLM based on:
\begin{enumerate*}[label=\roman*)]
    \item a post hoc participation in the 2025 edition of a state-of-the-art KGQA challenge (\S\ref{subsec:benchmark-general}),
    \item a set of questions from three of the most prevalent knowledge graphs within the field of bioinformatics (\S\ref{subsec:benchmark-bio}), and
    \item a live participation in the 2026 edition of the aforementioned KGQA challenge (\S\ref{subsec:benchmark-live}).
\end{enumerate*}

There exist two main categories of metrics for evaluating the performance of KGQA systems \cite{taghzouti2026t2s}:
\begin{enumerate*}[label=\roman*)]
\item \emph{Lexical Overlap Metrics}, which compute the textual similarity between two queries, and
\item \emph{Execution-Based Metrics}, which compute the result overlap between two queries.
\end{enumerate*}
In our evaluation we report the TREC F1 Score \cite{van2018pytrec_eval}, an execution-based metric also utilized by the aforementioned KGQA challenge.
Specifically, given the result set of a generated query ($G$), and the result set of the ground-truth query ($T$), the F1 Score is computed as:
\begin{center}
F1 Score = $\frac{2 \cdot |T \cap G|}{|T| + |G|}$    
\end{center}
If multiple variables are projected in a query, then the result set is flattened before computing the F1 Score.
To determine the overall F1 Score of a set of queries, we compute the unweighted average of the individual F1 Scores of each query.

\subsection{Post Hoc KGQA Evaluation}\label{subsec:benchmark-general}
As the first part of our evaluation, we employ the First International TEXT2SPARQL challenge\footnote{\url{https://text2sparql.aksw.org}}, which was colocated with the \emph{Extended Semantic Web Conference (ESWC) 2025} (named TEXT2SPARQL 2025 for the rest of the paper).
There are two subtasks in the challenge: 
\begin{enumerate*}[label=\roman*)]
\item a task with a well-known and well-documented KG (DBpedia), consisting of $100$ natural language questions in English and Spanish (respectively named \emph{DBpedia (EN)} and \emph{DBpedia (ES)} for the rest of the paper), and 
\item a task with an unknown and undocumented KG regarding a private corporation, consisting of $50$ questions in English (named \emph{Corporate} for the rest of the paper).
\end{enumerate*}
To assess our system, we followed the instructions of the challenge and deployed a dedicated API of our system, supporting two GET parameters, namely \emph{dataset} and \emph{question}.
%
Consequently, we utilized \emph{text2sparql-client}\footnote{\url{https://pypi.org/project/text2sparql-client}} to measure the performance of our system in terms of its \emph{F1 Score}.
Finally, to account for the variability of the underlying LLMs and ensure the reliability of our results, we performed all the experiments three times, consistently reporting the standard $95\%$ confidence interval.

\begin{figure}[t]
    \centering
    \includegraphics[width=0.8\textwidth]{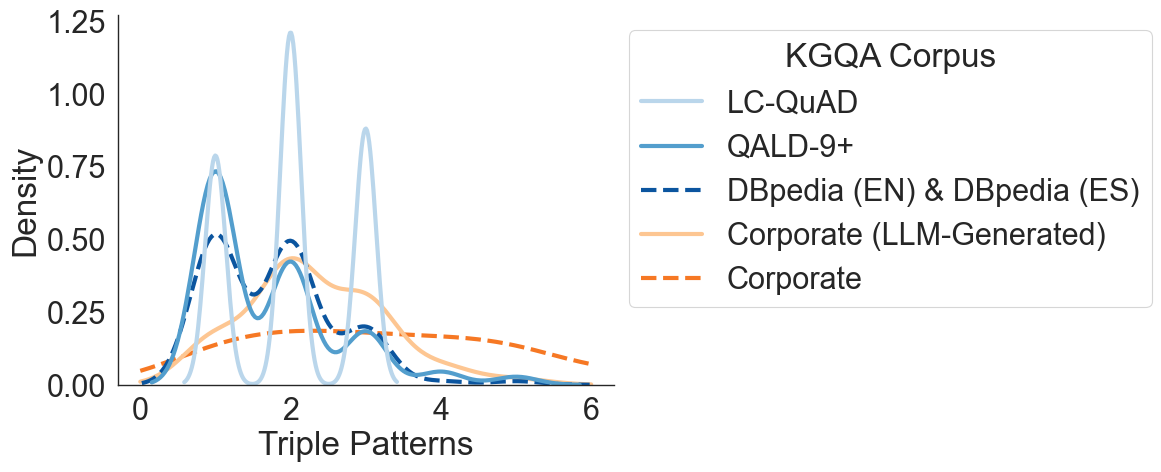}
    \caption{Kernel Density Estimate (KDE) of triple patterns of DBpedia-related (in blue) and Corporate-related (in orange) KGQA corpora. The dashed lines denote the evaluation corpora published by the TEXT2SPARQL 2025 challenge. We observe that the majority of the queries are considerably simple containing from one to three triple patterns.}
    \Description{Visualization of Kernel Density Estimate (KDE) of triple patterns for DBpedia and Corporate KGQA corpora, with dashed lines for TEXT2SPARQL challenge corpora.}
    \label{fig:dataset_details}
\end{figure}

\subsubsection{Provided Context}\label{subsec:context}
As mentioned above, there are two main types of context that we provide to SPARQL-LLM:
\begin{enumerate*}[label=\roman*)]
\item existing examples of pairs of natural language questions and SPARQL queries, and
\item schema information of the SPARQL endpoints.
\end{enumerate*}
%
%
For DBpedia, we use the example sets proposed by the challenge organizers, consisting of:
\begin{enumerate*}[label=\roman*)]
\item LC-QuAD \cite{DBLP:conf/semweb/TrivediMDL17}, a template-generated KGQA corpus, and
\item QALD-9+ \cite{DBLP:conf/semco/PerevalovDUB22}, an aggregate of nine editions of the KGQA corpus, QALD.
\end{enumerate*}
As we observe in Figure~\ref{fig:dataset_details}, almost all the DBpedia queries of the challenge contain between one and three triple patterns, similarly to the example queries of LC-QuAD and QALD-9+.

As for Corporate examples, since the KG is completely unknown and undocumented, and the provided ontology is relatively small (13 classes and 30 properties), we utilized an LLM, appropriately prompted to generate pairs of natural language questions and SPARQL queries\footnote{
You are an assistant that helps users formulate pairs of Natural Language Questions that correspond to SPARQL Queries.
Your role is to generate \{number\} example pairs based on the provided ontology consisting of 13 classes and 30 properties: \{ontology\}
}.
The LLM we utilized (DeepSeek-V3) was released before the publication of the test questions of the challenge; thus, it had no access to these questions during its training, and for fairness purposes, it was not included in the LLM selection experiment (Figure~\ref{fig:hyperparameters}).
%
%
Finally, as an additional verification step, we run the generated queries against the SPARQL endpoint of the Corporate KG and discard queries with no results.
Examples of such queries include: ``List products with missing price information'' and ``Show agents with multiple areas of expertise'', which cover the basic SPARQL functionality and the basic knowledge of the KG.

\begin{table}[t]
\caption{
LLM configuration of three sets of systems:
i) TEXT2SPARQL 2025: $ARUQULA$ is the winner of DBpedia (EN) and Corporate subtasks, while $mKGQAgent$ is the winner of the DBpedia (ES) subtask. To guarantee a fair comparison, all the $SPARQL-LLM$ variants use the same LLM Provider and configuration, and have knowledge cutoff before the release of the challenge;
ii) System Analysis (\S~\ref{sec:system_analysis}): different flagship LLMs are evaluated, having the same configuration and knowledge cutoff;
iii) TEXT2SPARQL 2026: $SPARQL-LLM$ is evaluated against four other systems, two of which were submitted with multiple variants to the challenge.
}
\label{table:LLM_configuration}
\begin{tabular}{r|ccccccc}
System & \shortstack{LLM\\Provider} & Version & Context & Temperature & \shortstack{Max\\Tokens} & Seed & \shortstack{Knowledge\\Cutoff} \\
\hline
\emph{TEXT2SPARQL 2025} \\
$SPARQL-LLM_{lg}$ & OpenAI & \texttt{GPT-4o} & 128K & 0 & 16K & 42 & 10/2023 \\
$SPARQL-LLM_{sm}$ & OpenAI & \texttt{GPT-4.1-mini} & 1.05M & 0 & 16K & 42 & 06/2024 \\
$SPARQL-LLM_{os}$ & OpenAI & \texttt{GPT-oss-120b} & 131K & 0 & 16K & 42 & 06/2024 \\
$ARUQULA$ \cite{Brei2025} & OpenAI & \texttt{GPT-4.1-mini} & 1.05M & 1 & 700 & - & 06/2024 \\
$mKGQAgent$ \cite{Perevalov2025} & OpenAI & \texttt{GPT-4o} & 128K & 0 & - & - & 10/2023 \\
\\ \emph{System Analysis} \\
$SPARQL-LLM$ & Mistral AI & \texttt{Mistral Large 2} & 131K & 0 & 16K & 42 & 05/2024 \\
$SPARQL-LLM$ & Qwen & \texttt{Qwen 2.5} & 32K & 0 & 16K & 42 & 03/2025 \\
$SPARQL-LLM$ & Anthropic & \texttt{Claude 3.5 Sonnet} & 200K & 0 & 16K & 42 & 04/2024 \\
\\ \emph{TEXT2SPARQL 2026} \\
$SPARQL-LLM$ & OpenAI & \texttt{GPT-5.2} & 400K & 0 & 16K & 42 & 08/2025 \\
$GRASP$ \cite{walter2025grasp} & OpenAI & \texttt{GPT-5.4-mini} & 400K & - & 8K & 22 & 08/2025 \\
$IRIS$ \cite{DBLP:conf/llms4ol/LatipovHM25} & Qwen & \texttt{Qwen 3.5} & 262K & - & - & - & 02/2026 \\
$StarCoder_{v1}$ \cite{Soru2025} & Anthropic & \texttt{Claude 4.5 Sonnet} & 200K & - & - & - & 01/2025 \\
$StarCoder_{v2}$ \cite{Soru2025} & Qwen & \texttt{Qwen 3.5} & 262K & - & - & - & 02/2026 \\
$ARUQULA_{v1}$ \cite{Brei2025} & OpenAI & \texttt{GPT-4.1-mini} & 1.05M & 1 & 700 & - & 06/2024 \\
$ARUQULA_{v2}$ \cite{Brei2025} & OpenAI & \texttt{GPT-4.1-mini} & 1.05M & 1 & 700 & - & 06/2024 \\
$ARUQULA_{v3}$ \cite{Brei2025} & OpenAI & \texttt{GPT-4.1-mini} & 1.05M & 1 & 700 & - & 06/2024 \\
\end{tabular}
\end{table}

\subsubsection{Systems}
We evaluate three variants of our system with the TEXT2SPARQL 2025 challenge:
\begin{enumerate*}[label=\roman*)]
\item $SPARQL-LLM_{lg}$, which is based on the flagship model \texttt{GPT-4o},
\item $SPARQL-LLM_{sm}$, which is based on the more compact model \texttt{GPT-4o-mini}, and
\item $SPARQL-LLM_{os}$, which is based on the open-source model \texttt{GPT-oss-120b}.
\end{enumerate*}
The knowledge cutoff for all the underlying LLMs was in $2023$ and $2024$ respectively, 
that is before the release of the challenge; therefore, it is impossible for them to have prior knowledge of the challenge datasets and questions, which would make the comparison with the other systems participating in the challenge unfair.
In the same note, since the challenge participants used \texttt{GPT} variants as their base LLM (Table~\ref{table:LLM_configuration}), we only evaluate variants of our system based on \texttt{GPT}.
An extended evaluation of other flagship LLMs is presented in Figure~\ref{fig:hyperparameters}.

\begin{figure}[t]
    \centering
    \includegraphics[width=0.7\textwidth]{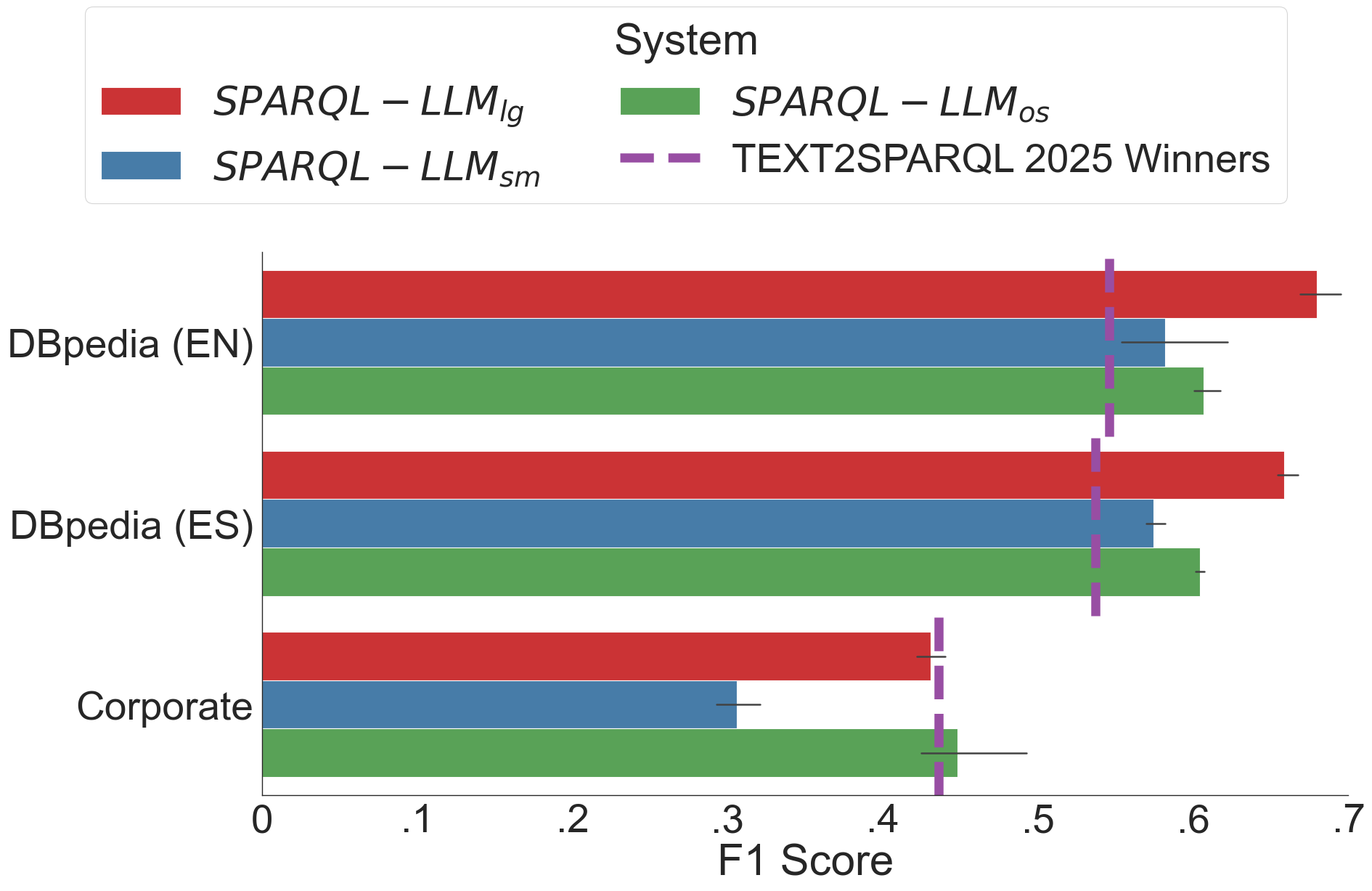}
    \caption{Performance evaluation of three variants of our system against the TEXT2SPARQL 2025 challenge winners, showcasing a substantial increase of $24\%$ in the \emph{F1 Score} for the DBpedia (EN) and DBpedia (ES) subtasks.}
    \Description{Bar chart showing F1 scores of systems competing in the challenge.}
    \label{fig:overall_results}
\end{figure}

\subsubsection{Results}
As we observe in the results (Figure~\ref{fig:overall_results}), SPARQL-LLM performs better in terms of \emph{F1 Score} across all three subtasks of the challenge.
For the DBpedia (EN) and DBpedia (ES) subtasks, the best performance is achieved with $SPARQL-LLM_{lg}$, showcasing a substantial increase of $24\%$ in the \emph{F1 Score}.
Interestingly, for the Corporate subtask, the best performance is achieved with $SPARQL-LLM_{os}$. In fact, $SPARQL-LLM_{os}$ is consistently better than the challenge winners across all three subtasks, which is encouraging considering the open-source nature of the underlying model. 

\begin{figure}[t]
    \centering
    \includegraphics[width=\textwidth]{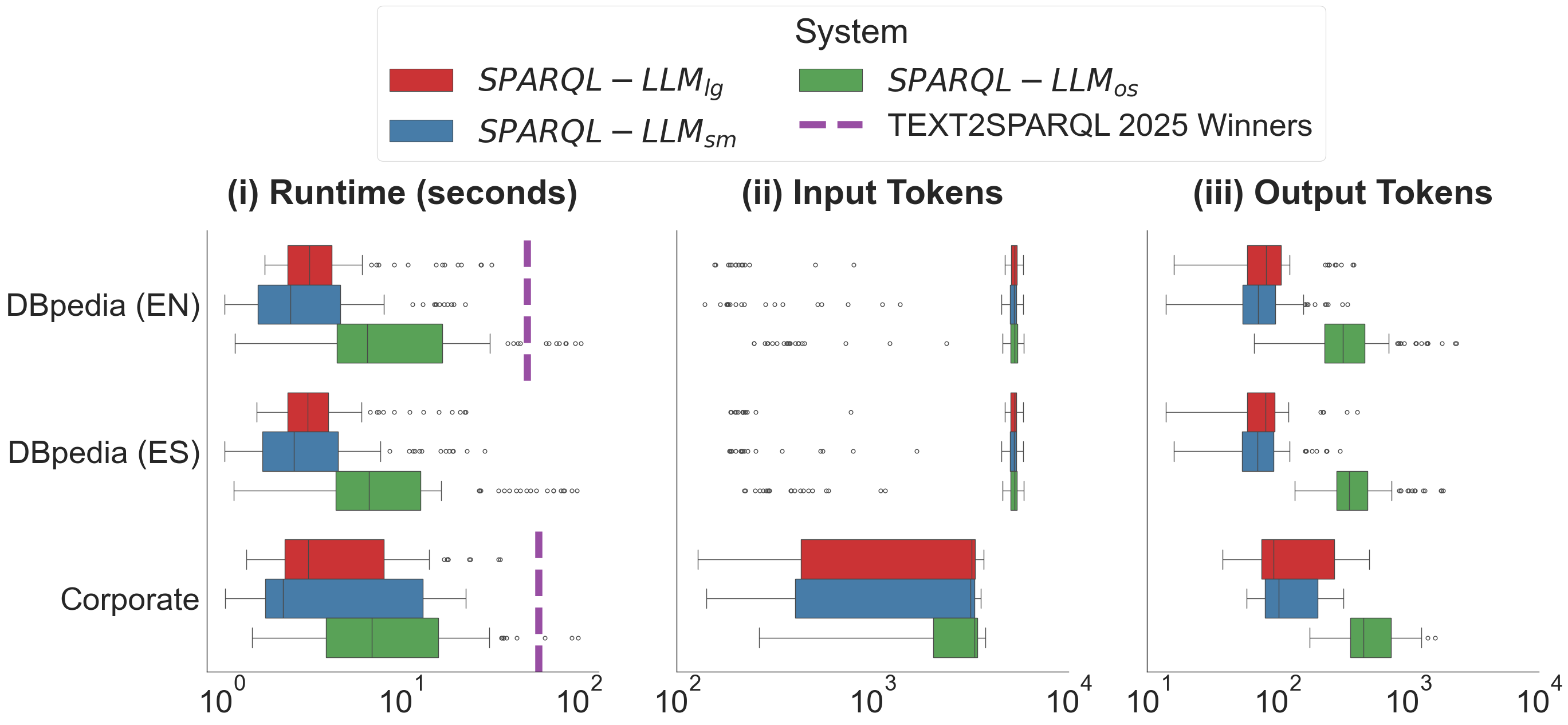}
    \caption{Cost analysis of the three versions of our system in terms of runtime and input/output tokens per question.}
    \Description{Bar chart showing the cost analysis of the three system versions by runtime and input/output tokens per question.}
    \label{fig:cost_results}
\end{figure}

\subsubsection{Cost Analysis}
An orthogonal evaluation to the performance in terms of accuracy (i.e., \emph{F1 Score}) involves assessing the performance of the systems regarding their cost, as indicated by their runtime and input/output LLM token usage.
This assessment is particularly crucial for systems deployed in production, with strict time and budget constraints.
As these extra evaluation dimensions were not systematically investigated by TEXT2SPARQL 2025, we highly recommend that they should be incorporated in future editions of the challenge and other text-to-SPARQL benchmarks.

Regarding the runtime (Figure~\ref{fig:cost_results} (i)), our system conforms to established usability engineering principles, which indicate that approximately $10$ seconds represents the threshold for sustaining a user’s focused attention within an interactive dialog \cite{nielsen1994usability}.
In more detail, $SPARQL-LLM_{sm}$ is, on average, slightly faster than $SPARQL-LLM_{lg}$, which can be explained by the size of the LLM, and $SPARQL-LLM_{os}$ seems to be the slowest variant, which is also expected since it is designed to run on off-the-shelf hardware without requiring dedicated large-scale GPU clusters.
Nonetheless, when compared to the average time to generate a SPARQL query, as reported by the TEXT2SPARQL 2025 challenge winner $ARUQULA$ \cite{Brei2025}, SPARQL-LLM is up to $27\times$ times faster\footnote{For Corporate, the median runtime of $SPARQL-LLM_{sm}$ is $2.2s$, while the corresponding runtime for $ARUQULA$ \cite{Brei2025} is $59.6s$.}.

Regarding the tokens (Figure~\ref{fig:cost_results} (ii) \& (iii)), given that the questions of DBpedia (EN) and DBpedia (ES), as well as, to a smaller extent, the questions of Corporate, have uniform lengths, and given also that the context provided to these questions has fixed size, we observe no significant variance in the number of input tokens.
Furthermore, we observe similar patterns in the output tokens of $SPARQL-LLM_{sm}$ and $SPARQL-LLM_{lg}$, with $SPARQL-LLM_{os}$ being a slightly more verbose variant.
Overall, with the current prices of the LLM tokens, the average cost per question for SPARQL-LLM is estimated to a maximum of $\$0.01$\footnote{The cost of the system variants is computed by averaging the input/output tokens and multiplying them by the cost reported by OpenRouter as of December 2025; $SPARQL-LLM_{lg}$: $\$0.01$, $SPARQL-LLM_{sm}$: $\$0.0007$, and $SPARQL-LLM_{os}$: $\$0.0004$.}.

\begin{figure}[t]
    \centering
    \includegraphics[width=\textwidth]{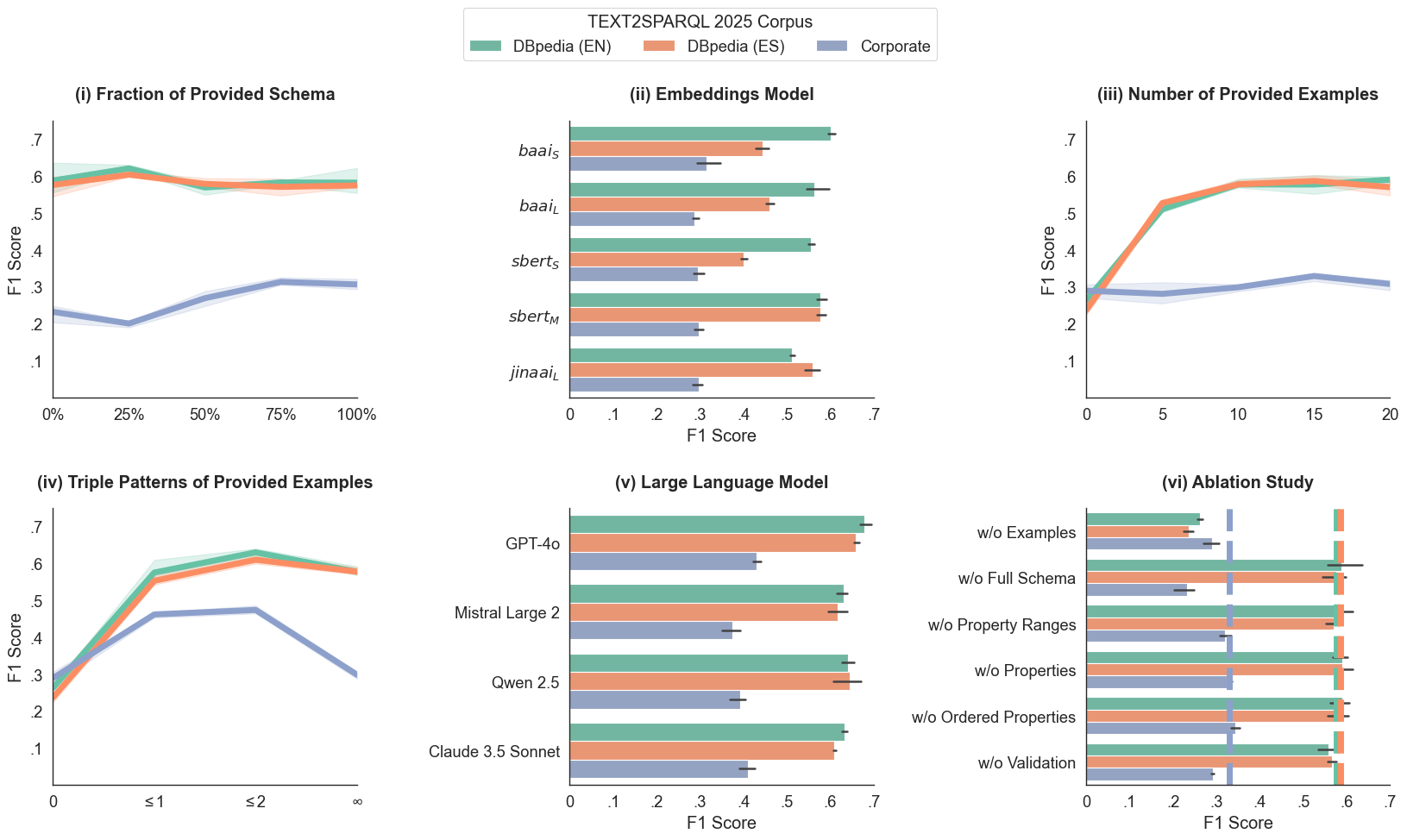}
    \caption{Hyperparameter fine-tuning and ablation study of SPARQL-LLM.}
    \Description{Results of hyperparameter tuning and ablation study for SPARQL-LLM are shown in the figure.}
    \label{fig:hyperparameters}
\end{figure}

\subsubsection{System Analysis}\label{sec:system_analysis}
To further investigate the performance of our system, we conduct a series of experiments in which we fine-tune several hyperparameters.
We note that this is a post-evaluation analysis intended to potentially serve as a reference for best practices when designing future enhancements of our system as well as other text-to-SPARQL systems.
Using $SPARQL-LLM_{sm}$ as our starting point, we focus on four hyperparameters:
\begin{enumerate*}[label=\roman*)]
\item the fraction of the data-aware schema of the endpoints that we index,
\item the embeddings model that we use for indexing,
\item the number of reference examples and classes provided to the underlying LLM, and
\item the LLM itself.
\end{enumerate*}
Furthermore, inspired by \cite{DBLP:conf/emnlp/LiuSTXZL24}, we conduct an ablation study by removing essential components of our system to measure their importance.

\paragraph{Fraction of Provided Schema}
First, we experiment with the frequency-based fraction of the schema that we provide to SPARQL-LLM (Figure~\ref{fig:hyperparameters} (i)).
To compute this fraction, we first compute a sparse class-property matrix with all the combinations of classes and properties of the schema, which we then 
sort on both dimensions respectively based on:
\begin{enumerate*}[label=\roman*)]
\item the number of instances of each class in the KG, and
\item the number of times that each property appears in the KG.
\end{enumerate*}
Consequently, we truncate the matrix to a specified fraction (e.g., to $25\%$), thus ensuring by construction that the resulting submatrix represents the $25\%$ most frequent fraction of the given schema.

As we observe in the results, for DBpedia, the optimal performance is achieved with only the $25\%$ most frequent fraction of the schema.
The latter is explained partially by the fact that there are many classes with the same name (e.g., \emph{Person} appears with three namespaces: \texttt{dbo:Person}, \texttt{schema:Person}, and \texttt{foaf:Person}), which makes the LLM struggle with selecting the right one.
Nonetheless, the fraction of the DBpedia schema provided to the LLM does not significantly affect the performance of the LLM, which suggests that the LLM is already (partially) aware of the DBpedia knowledge graph.
On the other hand, for the Corporate subtask, the more schema that is provided, the more accurate the LLM becomes, which is the expected behavior for such a completely-unknown-to-the-LLM knowledge graph.
Hence, for popular knowledge graphs, providing a smaller fraction of their schema does not affect the accuracy of the results, while reducing the context and, consequently, the per-question cost of the system.

\paragraph{Embeddings Model}
Regarding the embeddings model we use for indexing and retrieving reference SPARQL examples and classes, we evaluate the most prominent alternatives available directly in our vector database\footnote{\url{https://qdrant.github.io/fastembed/examples/Supported_Models}} (Figure~\ref{fig:hyperparameters} (ii)).
These models are: a small (\texttt{bge-small-en-v1.5}) version and a large (\texttt{bge-large-en-v1.5}) version of BGE~\cite{bge_embedding},
a small (\texttt{all-MiniLM-L6-v2}) and a medium and multilingual (\texttt{paraphrase-multilingual-mpnet-base-v2}) version of 
Sentence Transformers (SBERT) \cite{DBLP:conf/emnlp/ReimersG19}, and a large and multilingual Jina Embedding model (\texttt{jinaai/jina-embeddings-v3})~\cite{sturua2024jinaembeddingsv3multilingualembeddingstask}.
Although our comparison is not exhaustive, the selected models showcase remarkable performance in the related leaderboard\footnote{\url{https://huggingface.co/spaces/mteb/leaderboard}}.

The results of this experiment indicate that, for this particular task, the size of the embeddings does not significantly affect the performance of our system, with large models performing equally well or even worse than the respective small or medium models.
However, we observe that multilingual embedding models perform significantly better in DBpedia (ES), which is expected, as the retrieval is more precise than with the respective monolingual english embedding models.
Hence, when multilingual support is essential for a text-to-SPARQL system such as SPARQL-LLM, it is recommended to also employ a multilingual indexing and retrieval mechanism.

\paragraph{Number of Provided Examples}
Regarding the number of examples provided to the underlying LLM, we perform an experiment with 0 to 20 examples (Figure~\ref{fig:hyperparameters} (iii)).
The results for both DBpedia (EN) and DBpedia (ES) show that the performance of our system plateaus after 10 examples, while the context, and thus the per-question cost, increases.
Furthermore, the performance of SPARQL-LLM does not significantly change after introducing the LLM-generated examples described above.
Hence, given an adequate indexing and retrieval mechanism, 5 to 10 relevant and ideally non-overlapping examples are enough for such a system to perform efficiently.
%

\paragraph{Triple Patterns of Provided Examples}
Regarding the algebraic complexity of the examples provided to the underlying LLM, we perform an experiment with a varying number of triple patterns (Figure~\ref{fig:hyperparameters} (iv)).
As we observe, for the DBpedia (EN) and DBpedia (ES) subtasks, the performance of SPARQL-LLM increases when the example queries contain up to two triple patterns, after which it slightly declines.
This behavior is expected, as the majority of the test queries have a comparable level of complexity (Figure~\ref{fig:dataset_details}).
Thus, it seems that SPARQL-LLM achieves higher performance when the provided example queries have a complexity that is closely aligned with that of the queries it is required to generate.

For the Corporate subtask, we observe a substantial decline in performance when the example queries contain more than two triple patterns.
As discussed in \S\ref{subsec:context}, for this subtask, due to the lack of real-world examples, we generated synthetic examples with the assistance of another LLM.
The observed decline in performance suggests that the LLM-generated examples of lower structural complexity are likely to be semantically coherent and thus informative, whereas the more complex LLM-generated examples may introduce noise into the system.

\paragraph{Large Language Model}
Furthermore, we perform an experiment benchmarking the most prevalent and affordable flagship LLMs available on OpenRouter (Figure~\ref{fig:hyperparameters} (v)).
Specifically, we compare the following models: \texttt{GPT-4o}, \texttt{Mistral Large 2}, \texttt{Qwen 2.5}, and \texttt{Claude 3.5 Sonnet}.
Again, the list of models is not exhaustive, and the versions of the models date back prior to the release of the TEXT2SPARQL challenge to make the comparison less biased.

Overall, the results demonstrate that all the LLMs have similar performance, with \texttt{GPT-4o} showing slight superiority in this particular benchmark.
However, as the models evolve rapidly, without proper versioning and archiving, we note that this experiment may not be reproducible in the near future (e.g., the model \texttt{Gemini 1.5 Pro}, with a similar release date as the aforementioned models, is already unavailable for testing), and we recommend further investigations with the at-the-time most prevalent LLMs.
These experiments are not intended as a comprehensive benchmark but serve to illustrate how current general-purpose models perform within our framework.

\paragraph{Ablation Study}
Finally, we perform an ablation study by removing essential components of our system to measure their importance (Figure~\ref{fig:hyperparameters} (vi)).
As we observe in the results, SPARQL-LLM shows a significant decline in performance when examples are not supplied.
We further observe a decline when schema information about classes is not provided for the Corporate knowledge graph, while we do not observe such decline when property-related information (property ranges, properties, or their popularity-based ordering) is not provided; thus, class information seems to be the most useful segment of the schema.
As already mentioned, the supply of schema information has little-to-no effect on the DBpedia knowledge graph, as the underlying LLM seems to be already partially aware of it. 
Finally, we also observe a decline in performance when we remove the validation component of our system.
Overall, the essential components of SPARQL-LLM, in order of significance, are the examples, the schema class information, and the validation component.

    
    

\begin{table}[t]
\caption{Error analysis on the DBpedia (EN) questions: Errors categories (top), and examples of erroneous reference queries (bottom).
}
\label{table:error_analysis}

\scriptsize
\begin{tabular}{rrrr}
     \textbf{Error Description} & \textbf{Reference Query (Example)} & \textbf{Generated Query (Example)} & \textbf{Count} \\
     \hline
     Predicate/Class Namespace Mismatch & \texttt{dbo:origin} & \texttt{dbp:origin} & $20$ \\
     Erroneous Reference Query & (see below) & (see below) & 11 \\     
     Redundant Aggregate Variable Projection & \texttt{SELECT ?country WHERE} & SELECT ?country (COUNT(?city) AS ?cityCount) & $4$ \\
     Wrong Text Encoding & \texttt{dbr:Republic\_of\_Montenegro\_(1992–2006)} & \texttt{dbr:Republic\_of\_Montenegro\_(1992\%E2\%80\%932006)} & 3 \\
     Alternative Predicate Usage & \texttt{dbp:placeOfDeath} & \texttt{dbp:deathPlace} & 2 \\
     Empty Generated Query & \texttt{SELECT [\dots]} & - & 2 \\     
     Wrong Query Type & \texttt{ASK [\dots]} & \texttt{SELECT [\dots]} & 1 \\
     SPARQL Syntax: ORDER BY, LIMIT & - & - & 0 \\
      &  &  &  \\
\end{tabular}
\newline
\newline
\begin{tabular}[t]{b{0.45\textwidth}  b{0.45\textwidth}}
    \multicolumn{2}{c}{\textbf{Which country has the highest population?}}\\\\
    \cellcolor{gray!10}
    \begin{minipage}[t]{\linewidth}
    \begin{center}
        \textbf{Reference Query}
    \end{center}
    \begin{lstlisting}[language=SPARQL,basicstyle=\footnotesize\ttfamily]
    SELECT DISTINCT ?country WHERE {
        ?country dbo:populationTotal ?population .
    } ORDER BY DESC(?population) LIMIT 1
    \end{lstlisting}
    \end{minipage}
    &
    \begin{minipage}[t]{\linewidth}
    \begin{center}
        \textbf{Generated Query}
    \end{center}
    \begin{lstlisting}[language=SPARQL,basicstyle=\footnotesize\ttfamily]
    SELECT DISTINCT ?country  WHERE {
        ?country rdf:type dbo:Country ;
        dbo:populationTotal ?population
    } ORDER BY DESC(?population) LIMIT 1
    \end{lstlisting}
    \end{minipage}
    \\\\\\
    \multicolumn{2}{c}{\textbf{How many unique authors have written science fiction novels?}}\\\\
    \cellcolor{gray!10}
    \begin{minipage}[t]{\linewidth}
    \begin{center}
        \textbf{Reference Query}
    \end{center}
    \begin{lstlisting}[language=SPARQL,basicstyle=\footnotesize\ttfamily]
    SELECT DISTINCT (COUNT(?author) AS ?count) WHERE {
      ?x dbo:literaryGenre> dbr:Science_fiction .
      ?x dbo:author ?author .
    }
    \end{lstlisting}
    \end{minipage}
    &
    \begin{minipage}[t]{\linewidth}
    \begin{center}
        \textbf{Generated Query}
    \end{center}
    \begin{lstlisting}[language=SPARQL,basicstyle=\footnotesize\ttfamily]
    SELECT (COUNT(DISTINCT ?author) AS ?count) WHERE {
        ?x a dbo:Book .
        ?x dbo:literaryGenre dbr:Science_fiction .
        ?x dbo:author ?author .
    }
    \end{lstlisting}
    \end{minipage}
    \\\\\\
    \multicolumn{2}{c}{\textbf{When did the Apollo 11 mission land on the moon?}}\\\\
    \cellcolor{gray!10}    
    \begin{minipage}[t]{\linewidth}
    \begin{center}
        \textbf{Reference Query}
    \end{center}
    \begin{lstlisting}[language=SPARQL,basicstyle=\footnotesize\ttfamily]
    SELECT DISTINCT ?d  WHERE {
        dbr:Apollo_11 dbo:launchDate ?d .
    }
    \end{lstlisting}
    \end{minipage}
    &
    \begin{minipage}[t]{\linewidth}
    \begin{center}
        \textbf{Generated Query}
    \end{center}    
    \begin{lstlisting}[language=SPARQL,basicstyle=\footnotesize\ttfamily]
    SELECT ?date  WHERE {
        dbr:Apollo_11 dbo:landingDate ?date .
    }
    \end{lstlisting}
    \end{minipage}
\end{tabular}
\end{table}

\subsubsection{Error Analysis}
To complement our system analysis, we also performed a qualitative analysis of the errors of our system.
As we show in Table~\ref{table:error_analysis} (top), we discovered 7 different categories of errors; two of them, namely the \emph{Predicate/Class Namespace Mismatch} and the
\emph{Alternative Predicate Usage}, are caused because of the complex and often redundant schema of DBpedia, resulting in our system using predicates with the wrong namespace (e.g., \texttt{dbp:origin} instead of \texttt{dbo:origin}), or predicates with semantical duplicates (e.g., \texttt{dbp:deathPlace} instead of \texttt{dbp:placeOfDeath}).

Furthermore, we discovered the error category \emph{Redundant Aggregate Variable Projection}, where an aggregate variable used for sorting or grouping, is also projected in the final results (e.g., when sorting countries by their number of cities, this number is also projected).
Although the generated query may be semantically equivalent to the reference query, we observe a decline in the F1 Score, which occurs from the way the F1 Score is computed; as mentioned above, if multiple variables are projected in a query, then the result set is flattened before computing the F1 Score.

More errors include \emph{Wrong Text Encoding}, \emph{Wrong Query Type}, as well as \emph{Empty Generated Query}. 
Interestingly, no errors were observed when translating question components that require top-k aggregation (e.g., \emph{``Which country has the highest population?''}), into the correct SPARQL syntax using \emph{ORDER BY} and \emph{LIMIT}.

Surprisingly, in our analysis, we discovered $11$ cases of \emph{Erroneous Reference Query}, in which the query generated by our system was more accurate than the reference query. However, to ensure a fair comparison with the 2025 challenge winners, we did not correct the reference queries from the TEXT2SPARQL 2025 challenge.
The common element in most of these cases was that the generated query was binding variables to particular classes, which made the query more restrictive (but also more semantically accurate) than the reference ones.
In Table~\ref{table:error_analysis} (below), we see one such example for the question \emph{``Which country has the highest population?''}.
The result of the reference query is \emph{Mudikandam}, with a population of $9.2$ quintillion people, which is an artificially inflated number in the data dump provided by the challenge; nevertheless, \emph{Mudikandam} is not a \emph{Country} but a \emph{Village} in India.
In contrast, the result of the query generated by our system is \emph{Commonwealth of Nations}, which, in the data dump, is defined as a \emph{Country}; we highlight that neither of these answers is correct; however, this is a problem of the data dump itself and not of the generated query.  

Moreover, in the second example, for the question \emph{``How many unique authors have written science fiction novels?''}, we observe the same issue as above, where the variable representing a \emph{Book} is not bound to the appropriate class.
In addition, the reference query inverts the use of \emph{DISTINCT} and \emph{COUNT} in: \emph{SELECT DISTINCT (COUNT(?author) AS ?count)}; thus, it applies \emph{DISTINCT} to the aggregated count rather than to the variable being counted. 
As we observe, the generated query uses the correct formulation: \emph{SELECT (COUNT(DISTINCT ?author) AS ?count)}.

Finally, in the third example, for the question \emph{``When did the Apollo 11 mission land on the moon?''}, the reference query uses the wrong predicate \texttt{dbo:launchDate}, while the generated query uses the correct predicate \texttt{dbo:landingDate}.

\begin{figure}[t]
    \centering
    \includegraphics[width=0.4\textwidth]{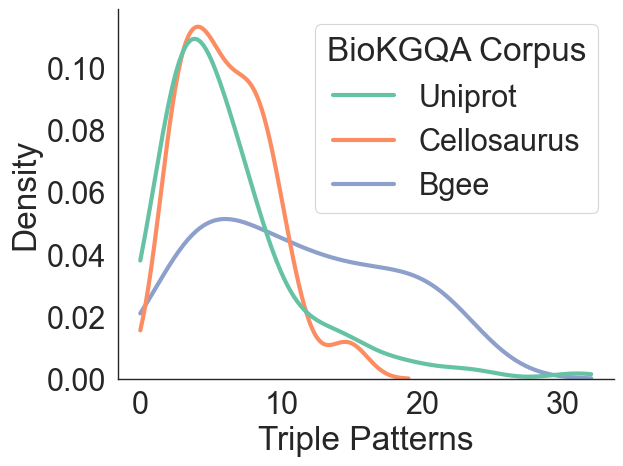}
    \caption{Kernel Density Estimate (KDE) of triple patterns of the three of the most prevalent bioinformatics corpora. We observe that all the BioKGQA queries contain up to $10\times$ more triple patterns than the classic KGQA queries (Figure~\ref{fig:dataset_details}).}
    \Description{Visualization of Kernel Density Estimate (KDE) of triple patterns for bioinformatics corpora.}
    \label{fig:bio_dataset_details}
\end{figure}

\subsection{BioKGQA Evaluation}\label{subsec:benchmark-bio}
We also evaluate SPARQL-LLM using three of the most prevalent knowledge graphs in the field of bioinformatics, namely Uniprot \cite{10.1093/nar/gkae1010}, Cellosaurus \cite{bairoch2018cellosaurus}, and Bgee \cite{DBLP:journals/nar/BastianRNCCFMPL21}.
We highlight that all these knowledge graphs are accessible through different public SPARQL endpoints and already include lightweight metadata regarding:  
\begin{enumerate*}[label=\roman*)]
\item example pairs of questions in natural language and related SPARQL queries expressed using SHACL vocabulary terms \cite{DBLP:journals/corr/abs-2410-06010}, and
\item data-aware schema information,  
\end{enumerate*} expressed using ShEx. However, as outlined in Section~\ref{sec:indexing_component}, this metadata can also be automatically generated from scratch for any given (potentially domain-specific) knowledge graph.

The number of example queries for the three knowledge graphs is: $123$ for Uniprot, $73$ for Cellosaurus, and $28$ for Bgee; comparable to the respective number of queries of TEXT2SPARQL 2025.
Moreover, as we observe (Figure~\ref{fig:bio_dataset_details}), these queries are significantly more complex than the queries of TEXT2SPARQL 2025; indeed, some of them contain $10\times$ more triple patterns. 
%
Thus, reconstructing such queries from their natural language descriptions is a very challenging task, even for domain-experts who are ``fluent'' in SPARQL.

To evaluate our system, we follow a standard 
3-fold cross-validation methodology, where we provide an \emph{example set} of queries to SPARQL-LLM, and evaluate its performance on a respective \emph{evaluation set} of queries.
It is worth mentioning that in order to perform cross-validation:
\begin{enumerate*}[label=\roman*)]
\item we assume there is no correlation among the examples; i.e., each example is semantically distant from all the others, both in its textual description and its SPARQL representation, and
\item we exclude queries that return zero results as there is no way to evaluate them; we employ the same (result-based) evaluation as in the TEXT2SPARQL 2025 challenge and report accordingly the F1 score.
\end{enumerate*}
As this is an endpoint-agnostic methodology, we strongly recommend that future evaluations of text-to-SPARQL systems consider it as an alternative to the well-studied use-cases based on the DBpedia, Wikidata, and DBLP knowledge graphs (details in Section~\ref{sec:related_work}).

\begin{figure}[t]
    \centering
    \includegraphics[width=0.8\textwidth]{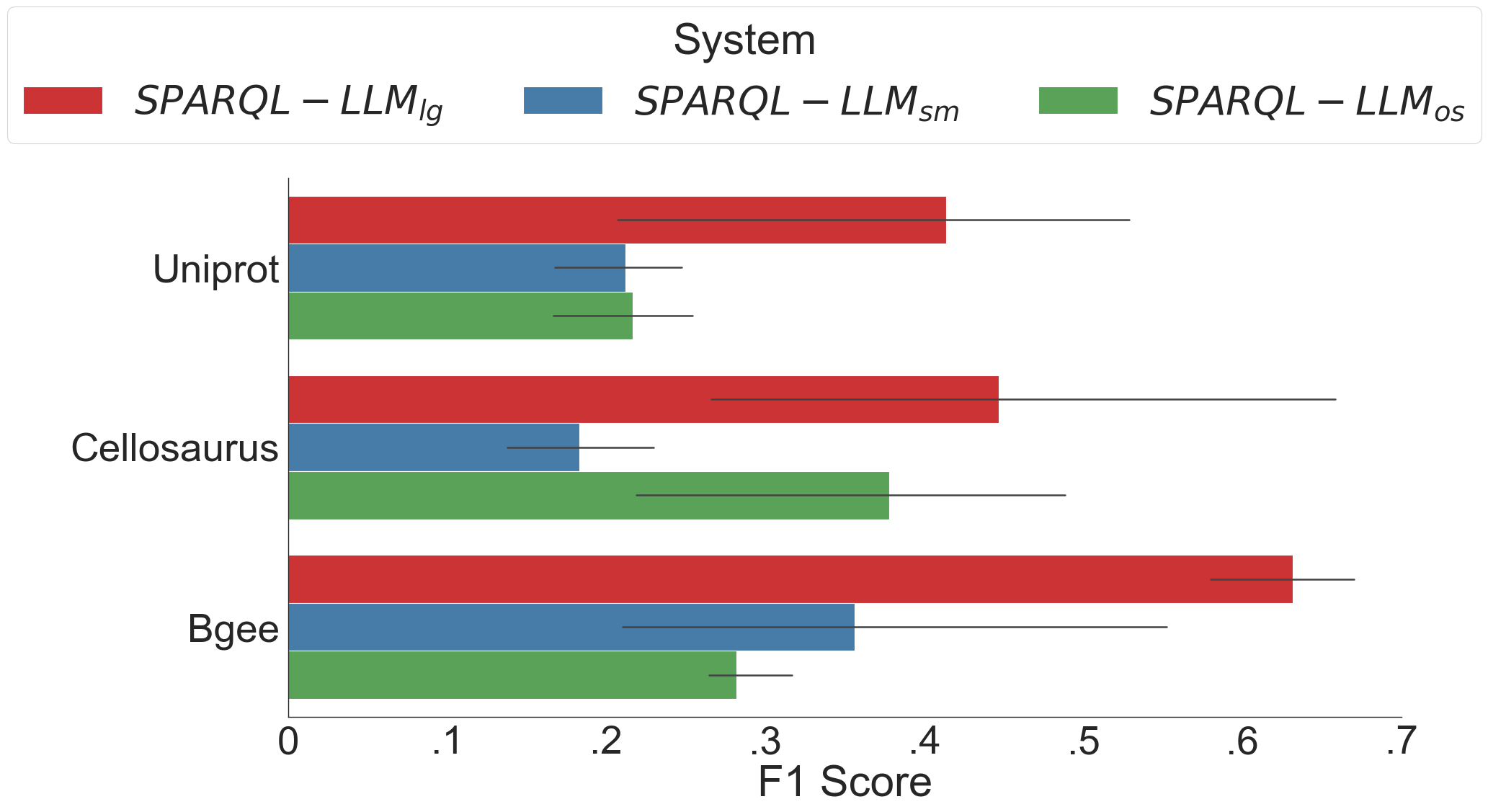}
    \caption{Results with BioKGQA evaluation queries.}
    \Description{Visualization of results of BioKGQA benchmark queries.}
    \label{fig:bio_results}
\end{figure}

As we observe in the results (Figure~\ref{fig:bio_results}), the performance of SPARQL-LLM slightly deteriorates in the BioKGQA evaluation compared to its performance in the standard KGQA evaluation, while $SPARQL-LLM_{lg}$ remains the most effective variant of our system.
Since, as explained above, reconstructing such complex queries is a challenging task even for domain-experts, we acknowledge the encouraging prospects shown by these results.

GRASP \cite{walter2025grasp} is also a recent system that is evaluated with a subset of the BioKGQA corpora, specifically with UniProt. However, the results of the two systems are not directly comparable since:
\begin{enumerate*}[label=\roman*)]
\item GRASP reports an \emph{F1-score} of $0.207$ over $50$ out of $120$ queries that were randomly selected from the UniProt website on May~7,~2025, while
SPARQL-LLM reports an \emph{F1-score} of $0.414$ over a total of 126 queries acquired from the UniProt website on November~17,~2025\footnote{\url{https://github.com/sib-swiss/sparql-examples/tree/d660276cb411c8e06216e167ea336a42fe004777/examples/UniProt}}, and
\item GRASP uses a different data dump consisting of only ``4M entities'' as per the indexed data described in the paper, whereas SPARQL-LLM uses the live public endpoint\footnote{\url{https://sparql.uniprot.org}} consisting of billions of entities.
\end{enumerate*}
The latter also demonstrates that our system scales well even with very large triplestores, since it only needs to index lightweight metadata and does not rely on performing on-the-fly searches during the SPARQL query generation.

\begin{table}[t]
\caption{
Results of the TEXT2SPARQL 2026 challenge, reported as F1 scores. Across both subtasks and all three languages, $SPARQL-LLM$ showcases a substantial improvement of up to $59\%$ in F1 score over the second-best system ($ARUQULA_{v1}$).
}
\label{table:results_2026}
\begin{tabular}{r|cccc}
& \multicolumn{2}{c}{DBpedia} & \multicolumn{2}{c}{Corporate} \\
System & English & Spanish & English & German \\
\hline
$SPARQL-LLM$                           & \textbf{.7192} & \textbf{.7810} & \textbf{.8510} & \textbf{.8104} \\
$GRASP$ \cite{walter2025grasp}                & .3296 & .3288 & .5262 & .4892 \\
$IRIS$ \cite{DBLP:conf/llms4ol/LatipovHM25}   & .2897 & .2914 & .2896 & .1460 \\
$StarCoder_{v1}$ \cite{Soru2025}              & .3235 & .3730 & .3215 & .3510 \\
$StarCoder_{v2}$ \cite{Soru2025}              & .3129 & .2361 & .1724 & .2390 \\
$ARUQULA_{v1}$ \cite{Brei2025}                & .6142 & .4904 & .6437 & .6230 \\
$ARUQULA_{v2}$ \cite{Brei2025}                & .2539 & .2200 & .1937 & .1148 \\
$ARUQULA_{v3}$ \cite{Brei2025}                & .0000 & .0000 & .5279 & .3827 \\
\end{tabular}
\end{table}

\subsection{Live KGQA Evaluation}\label{subsec:benchmark-live}

Based on our experience with the post hoc KGQA Evaluation as well as our system analysis and ablation study, we decided to participate in the 2026 edition of the TEXT2SPARQL challenge, which is colocated with the \emph{Extended Semantic Web Conference (ESWC) 2026} (named TEXT2SPARQL 2026 for the rest of the paper).
Similarly to the previous edition of the challenge, there are two subtasks in TEXT2SPARQL 2026: 
\begin{enumerate*}[label=\roman*)]
\item a task with DBpedia, consisting of $50$ natural language questions in English and Spanish, and 
\item a task with a KG regarding a private corporation, consisting of $50$ questions in English and German.
\end{enumerate*}
To participate in the challenge, we followed the instructions of the organizers and deployed a dedicated API of our system, supporting two GET parameters, namely \emph{dataset} and \emph{question}.

For $SPARQL-LLM$, we kept exactly the same configuration as described in \S\ref{subsec:benchmark-general}; the only difference is that we used \texttt{GPT-5.2} as the base LLM for our system (details in Table~\ref{table:LLM_configuration}). Furthermore, we provided the same context as described in \S\ref{subsec:context}, with the addition of the test corpora of TEXT2SPARQL 2025 (i.e., DBpedia (EN), DBpedia (ES), and Corporate).

\paragraph{Results}
As shown in Table~\ref{table:results_2026}, $SPARQL-LLM$ is evaluated against four other systems, two of which were submitted with multiple variants to the challenge. 
The configuration details for these systems are described in Table~\ref{table:LLM_configuration}. 
Across both subtasks and all three languages, $SPARQL-LLM$ showcases a substantial improvement in F1 score over the second-best system as ranked by the official challenge. 
More specifically, $SPARQL-LLM$ achieves a $17\%$ improvement in F1 score on DBpedia (English), $59\%$ on DBpedia (Spanish), $32\%$ on Corporate (English), and $30\%$ on Corporate (German).
Furthermore, as we observe in the detailed official results of the challenge\footnote{\url{https://aksw.github.io/text2sparql.aksw.org/2026/results}}, $SPARQL-LLM$ outperforms the other systems across all evaluation metrics, i.e., Precision, Recall, and NDCG (Normalized Discounted Cumulative Gain), demonstrating the robustness of our system and the consistency of the overall results.
Finally, the total cost of all the LLM calls made by $SPARQL$-$LLM$ for TEXT2SPARQL 2026 was less than $\$3$, confirming that it is not only the most performant system but also a system with minimal operational cost.

%% file: 05-conclusion.tex
\section{Conclusion}\label{sec:conclusion}

We presented SPARQL-LLM, an open-source, triplestore-agnostic approach designed to generate SPARQL queries from natural language.
By leveraging lightweight metadata from SPARQL endpoints and constructing sophisticated LLM prompts enriched with this contextual information, SPARQL-LLM outperforms existing approaches that rely on iterative interactions between traversing a knowledge graph and refining their LLM prompts.
Hence, it provides a solution with the highest accuracy, lowest latency, and minimal operational costs compared to state-of-the-art systems.
SPARQL-LLM is deployed in production, helping SPARQL experts  and non-experts generate and execute complex queries against real-world decentralized knowledge graphs.

\paragraph{Limitations}
Limitations of this work include the evaluation of our system with KGQA corpora: 
\begin{enumerate*}[label=\roman*)]
\item in other high- and low-resource languages,
\item 
in additional domains beyond encyclopedic, corporate, and bioinformatics,
\item systematically covering the full SPARQL functionality, and
\item 
systematically assessing the quality and semantic relevance of the provided examples, including the coverage of query patterns, and the potential effect of noisy examples.
\end{enumerate*}
As our evaluation is constrained by the KGQA benchmarks made available by the research community, we advocate for the development of more diverse and continuously-evolving such benchmarks.

We also acknowledge the difficulty of fully isolating the contribution of the provided context from the prior knowledge of the LLMs; an experiment using LLMs with no prior exposure to SPARQL or programming languages would be ideal but is currently infeasible because the high-end LLMs we evaluate are trained on closed corpora.
Hence, we encourage the research community to publish LLMs with open training corpora, such as Apertus \cite{DBLP:conf/acl/Hernandez-CanoH26}, as this would enable more controlled analyses of model capabilities
and more formal guarantees of query-generation correctness.

\paragraph{Future Work}
Future work concentrates, among others, on:
\begin{enumerate*}[label=\roman*)]
\item extending SPARQL-LLM to systematically handle more complex federated queries across knowledge graphs,
\item
comparing different languages for schema representation, such as ShEx, SHACL RDF, SHACL Compact Syntax, and OWL, to evaluate their impact on token usage and query-generation accuracy,
\item 
experimenting with context-retrieval strategies alternative to the standard embeddings-based (e.g., keyword-based),
\item 
experimenting with context-retrieval strategies that include schema elements beyond classes such as predicates and named entities,
\item performing a qualitative study focusing on user experience,
\item deploying SPARQL-LLM as a Model Context Protocol (MCP) service to directly integrate out-of-the-box with existing agent harnesses (e.g. OpenCode, Goose),
\item extending the query generation to other data accessing methods (e.g., FastAPI queries), and
\item generating an end-to-end workflow from accessing, to wrangling, and analyzing any given data. 
\end{enumerate*}



\section*{Acknowledgments}

This work was funded by the SNSF MetaboLinkAI project, grant: 10.002.786 and by SERI via the HORIZON-INFRA-2024-EOSC-01-05 project EOSC Data Commons under grant agreement number 101188179. 

\section*{Declaration on Generative AI}
During the preparation of this paper, the authors used Generative AI tools such as Claude and GPT for grammar, spelling, paraphrasing, and stylistic edits. After using these tools, the authors reviewed and edited the content as needed; hence, the authors take full responsibility of the content of this paper.